\documentclass[runningheads]{llncs}
\usepackage{graphicx}
\usepackage{todonotes}
\usepackage{hyperref}
\usepackage[multiple]{footmisc}
\begin{document}
\title{The Impact of Process Complexity on Process Performance: A Study using Event Log Data}
\titlerunning{The Impact of Process Complexity on Process Performance}
%
\author{Maxim Vidgof \inst{1}\orcidID{0000-0003-2394-2247} \and
Bastian Wurm\inst{2}\orcidID{0000-0002-1002-5397} \and
Jan Mendling\inst{1,3,4}\orcidID{0000-0002-7260-524X}\thanks{The research by Jan Mendling was supported by the Einstein Foundation Berlin under grant EPP-2019-524 and by the German Federal Ministry of Education and Research under grant 16DII133.}}
\authorrunning{M. Vidgof et al.}
%
\institute{Wirtschaftsuniversit{\"a}t Wien, Welthandelsplatz 1, 1020 Vienna, Austria
\email{maxim.vidgof@wu.ac.at}
\and LMU Munich School of Management, Ludwigstrasse 28, 80539 Munich, Germany
\email{bastian.wurm@lmu.de}
\and Humboldt-Universit{\"a}t zu Berlin, Unter den Linden 6, 10099 Berlin, Germany
\email{jan.mendling@hu-berlin.de}
\and
Weizenbaum Institute, Hardenbergstraße 32, 10623 Berlin, Germany
}
\maketitle              
\begin{abstract}
Complexity is an important characteristic of any business process. The key assumption of much research in Business Process Management is that process complexity has a negative impact on process performance. So far, behavioral studies have measured complexity based on the perception of process stakeholders. The aim of this study is to investigate if such a connection can be supported based on the analysis of event log data. To do so, we employ a set of 38 metrics that capture different dimensions of process complexity. We use these metrics to build various regression models that explain process performance in terms of throughput time. We find that process complexity as captured in event logs explains the throughput time of process executions to a considerable extent, with the respective R-squared reaching up to 0.96. Our study offers implications for empirical research on process performance and can serve as a toolbox for practitioners. 
\end{abstract}
\section{Introduction}\label{sec:introduction}
Business processes management (BPM) provides various analysis techniques for improving the performance of business processes (see, for example, \cite{DBLP:books/sp/DumasRMR18}). Several of these techniques support the identification of root causes behind performance issues of a process. Some studies have pointed to the connection between process complexity as a root cause of bad process performance. More specifically, it has been established that standardized business processes are connected with better process performance~\cite{munstermann2010performance} and outsourcing success~\cite{wullenweber2008impact}. For this reason, high business process complexity is often a motivation for business process redesign initiatives~\cite{gunasekaran1997role}, but also a challenge for standardization efforts~\cite{schafermeyer2012impact}.

However, these studies 
largely build on perceptual measures, which entails at least three key issues. First, such measures require specific attention in order to meet potential validity concerns~\cite{ketokivi2004perceptual}. Second, perceptual differences exist along the organizational hierarchy. The so-called hierarchical erosion effect states that perceptions become less favourable towards the lower levels of the hierarchy~\cite{gibson2019hierarchical}. Third, a study based on perceptual measures is often restricted to an observation at only a single point in time. All of this raises the question to which extent a more precise investigation of the connection between process complexity and performance is possible. 

In this paper, we address this research problem. To this end, we utilize available event log dataset to a) calculate complexity measures and b) throughput time as a performance measure over different time windows. The connection between complexity and performance is then investigated by means of statistical regression.
Our results suggest that process complexity is closely connected to throughput time, but also dependent on idiosyncratic factors. We discuss implications of this finding for research and practice.

The remainder of the paper is structured as follows. Section~\ref{sec:background} introduces complexity metrics and the related concepts. Section~\ref{sec:approach} presents our approach for the calculation of process complexity, throughput time, and the creation of our statistical models.  Section~\ref{sec:results} presents our results and showcases the best statistical models. Section~\ref{sec:discussion} provides the discussion of the results and points to avenues for future research. Finally, Section~\ref{sec:conclusion} concludes with a summary. 


\section{Background}\label{sec:background}
In this section, we discuss the background against which we position our work. First, we summarize related work on process complexity. Second, we outline research on process performance and the role that complexity plays for it. 


\subsection{Process Complexity}
\label{sec:complexity}

In BPM research, process complexity has often been approached from a process model perspective. Most notably is the work by Mendling on the relationship between process model complexity and error probability \cite{DBLP:books/sp/Mendling2008,mendling2010seven,mendling2012thresholds}. Recently, various metrics for complexity based on event logs have been defined, partially inspired by work in neighboring disciplines, such as organization science. These measures can be used to quantify different aspects of business processes complexity that are visible from event log data. The various measures can be organized in five categories as presented in Table \ref{tab:measures}.

The first category encompasses measures pertaining to the \textit{size} of a given event log. These measures count properties of an event log, such as the number of \textit{events}, \textit{sequences}, and minimum, average, and maximum \textit{sequence length}~\cite{christianguenther-phd}.  

The second category contains measures capturing the variation of process behavior as documented in the event log. Many of the measures in this category build on a transition matrix that is derived based on the directly-followed relations as captured in the event log \cite{DBLP:journals/isci/AugustoMVW22}. Pentland et al. \cite{DBLP:journals/misq/PentlandLKH20} operationalize process complexity as the number of \textit{acyclic paths} provided by the transition matrix. Closely related to this is the measure by H{\ae}rem et al. \cite{haerem2015task}, who measure complexity as the number of \textit{ties}, i.e. directly-follows relations, over all distinct \textit{sequences}. Further measures that depict variation are Pentland's \cite{pentland2003conceptualizing} approach to compress an event log based on the Lempel-Ziv algorithm as well as the absolute and relative number of \textit{unique sequences} \cite{ProcessMiningBook} contained in an event log.

The third category includes measures that are based on different notions of distance \cite{DBLP:journals/isci/AugustoMVW22}. G\"unther \cite{christianguenther-phd} suggests a measure of \textit{affinity} of two event sequences, capturing the extent to which directly-follow relations of the sequences overlap. His \textit{average affinity} measure calculates the mean of the pair-wise affinity over all sequences in the event log \cite{christianguenther-phd}. This measure is similar to Pentland's \cite{pentland2003conceptualizing} \textit{deviation from random} of the transition matrix. Pentland \cite{pentland2003conceptualizing} further proposes average \textit{edit distance} between event sequences based on optimal matching \cite{cornwell2015social}. 

The fourth category of measures builds on graph entropy and has been recently proposed by Augusto et al.~\cite{DBLP:journals/isci/AugustoMVW22}. They distinguish between measures for sequence and variant entropy of an event log. Additionally, they suggest that each of the measures can be normalized to take a value between 0 and 1. 
We refer to these measures as simple entropy measures. 

Fifth, the measures by \cite{DBLP:journals/isci/AugustoMVW22} have been extended beyond the control flow to incorporate data variety~\cite{DBLP:conf/icpm/VidgofM22}. In contrast to the simple entropy measures, we refer to this class of measures as \textit{enriched entropy} measures.



\begin{table}[h]
\centering
\caption{Complexity Measures for Business Processes based on Event Logs (adapted from \cite{DBLP:journals/isci/AugustoMVW22})}
    \begin{tabular}{r|l|c}
    \hline
    \multicolumn{1}{l|}{Category} & Measure & Reference \\
    \hline
    \multicolumn{1}{l|}{Size} & Number of Events  & \cite{christianguenther-phd} \\
          & Number of Event Types & \cite{christianguenther-phd} \\
          & Number of Sequences & \cite{christianguenther-phd} \\
          & Minimum, Average, Maximum Sequence Length & \cite{ProcessMiningBook} \\
    \hline
    \multicolumn{1}{l|}{Variation} & Number of Acyclic Paths in Transition Matrix & \cite{DBLP:journals/misq/PentlandLKH20} \\
          & Number of Ties in Transition Matrix & \cite{haerem2015task} \\
          & Lempel-Ziv Complexity & \cite{pentland2003conceptualizing} \\
          & Number and Percentage of Unique Sequences & \cite{ProcessMiningBook} \\
    \hline
    \multicolumn{1}{l|}{Distance} & Average Affinity & \cite{christianguenther-phd} \\
          & Deviation from Random & \cite{pentland2003conceptualizing} \\
          & Average Edit Distance & \cite{pentland2003conceptualizing} \\
    \hline
    \multicolumn{1}{l|}{Simple Entropy} & Sequence entropy & \cite{DBLP:journals/isci/AugustoMVW22} \\
            & Variant entropy & \cite{DBLP:journals/isci/AugustoMVW22} \\
            & Normalized sequence entropy & \cite{DBLP:journals/isci/AugustoMVW22} \\
            & Normalized variant entropy & \cite{DBLP:journals/isci/AugustoMVW22} \\
    \hline
     \multicolumn{1}{l|}{Enriched Entropy } & Enriched Sequence entropy & \cite{10.1007/978-3-031-27815-0_7} \\
            & Enriched Variant entropy & \cite{10.1007/978-3-031-27815-0_7} \\
            & Enriched Normalized sequence entropy & \cite{10.1007/978-3-031-27815-0_7} \\
            & Enriched Normalized variant entropy & \cite{10.1007/978-3-031-27815-0_7} \\
    \hline
    \end{tabular}%

  \label{tab:measures}%
\end{table}%

Several of these measures have been applied to study an increasing breadth of research problems. The above named study by Augusto et al. \cite{DBLP:journals/isci/AugustoMVW22}, for example, investigated the influence of process complexity on the quality of process models derived from event log data. They find that process complexity is negatively correlated with the quality of discovered process models. Thus, the more complex the event log, the poorer will be the model discovered by process mining algorithms. Importantly, different discovery algorithms are more sensitive to certain complexity measures than others \cite{DBLP:journals/isci/AugustoMVW22}.

There are also some behavioral studies that examine how process complexity changes over time. Pentland et al. \cite{DBLP:journals/misq/PentlandLKH20} simulate how process complexity changes over time. They find that organizational processes undergo different phases of process complexity. At the initiation of their simulation, processes exhibit low levels of complexity. After several iterations of the simulation, process complexity suddenly sharply increases, leading to \textit{bursts} of complexity. Afterwards, complexity again decreases resulting in limited but ongoing variation in the process. Further, Wurm et al. \cite{wurm2021measuring} investigate process complexity in the Purchase-to-Pay and Order-to-Cash processes of a multinational enterprise. While they find that process complexity changes continuously, they do not find any indication for sudden bursts of complexity in the examined processes.

Importantly, both studies \cite{DBLP:journals/misq/PentlandLKH20,wurm2021measuring} rest on measures that are not precise \cite{DBLP:journals/isci/AugustoMVW22}. As shown in \cite{DBLP:journals/isci/AugustoMVW22}, corner cases can be identified that illustrate that the measures used tend to overestimate the actual complexity of a process.

\subsection{Process Performance}

The literature suggests a clear link between process complexity and process performance. Empirical studies indicate that the standardization of business processes ultimately leads to better process performance~\cite{munstermann2010performance} and outsourcing success~\cite{wullenweber2008impact}. In particular, Münstermann et al. \cite{munstermann2010performance} have found that process standardization is positively associated with different process performance dimensions, such as process time, cost, and quality. By means of standardization, organizations aim to reduce process complexity, i.e. the number of ways that a process can be performed \cite{pentland2020process,wurm2018development}.

At a second look, however, the relationship between process complexity and process performance is not that clear-cut. Detailed findings by Münstermann et al. \cite{munstermann2010performance} show that the effect of standardization is conditional to the industry and type of process in question. Specifically, they find that process standardization only significantly influences process performance in the service industry and for companies that can be classified as analyzers \cite{munstermann2010performance}.

Furthermore, studies that measure the complexity of processes and their corresponding performance rely on perceptual measures that can cause several important validity issues. First, there are important validity concerns that need to be taken into account when developing perceptual measures for organizational and process performance \cite{ketokivi2004perceptual}. Second, there are perceptual differences that need to be considered when interpreting results from perceptual measures. For example, the hierarchical erosion effect \cite{gibson2019hierarchical} describes that perceptions at lower levels of an organization's hierarchy tend to be less favourable. Similarly, Pentland \cite{pentland2003conceptualizing} shows that process stakeholders' perception and actual enactment of process variation diverge substantially. Third, the use of perceptional measures to determine changes in properties of business processes is inefficient. In order to assess the effect of an improvement initiative on process performance, one would have to survey process stakeholders again and again. For example, to assess the success of a standardization initiative, a company would have to survey process stakeholders at least twice: prior to the initiative and after the initiative. Thus, studies based on perceptual measures are often restricted to data that is collected at a single point in time and only provide a static perspective of processes and their performance.

In light of these limitations, several authors have proposed to use process mining to move from opinion-based to evidence-based measures for business processes \cite{cho2017evaluating,DBLP:journals/is/CappielloCPF22,recker2021scientific,DBLP:conf/hicss/GrisoldWMB20}. In this regard, the studies by \cite{DBLP:journals/is/CappielloCPF22},\cite{cho2017evaluating} are the first to define measurability of process performance indicators and evaluate process redesign best practices based on event logs, respectively.

In the following, we develop an evidence-based and time-sensitive measure for process complexity based on the recent work by \cite{DBLP:journals/isci/AugustoMVW22}. As this measure is based on graph entropy it is precise and allows researchers and process managers to quantify process complexity in a comprehensive way at any given point in time. In addition, it allows to continuously  monitor how process complexity changes over time.
We further evaluate the measure by applying it to a set of event logs from the Business Process Intelligence (BPI) Challenge allowing us to closely examine the relationship between process complexity and process performance.



\section{Approach}\label{sec:approach}

In this section, we describe our approach. First, we introduce the notion of forgetting that allows us to weight events in the event log differently based on their time of occurrence. Then, we prepare the dataset by splitting it into time periods, performing complexity and performance measurements as well as filtering out the outliers. Then, we automatically build regression models and systematically reduce the number of variables in them.
All computations were performed on a laptop with Intel\textregistered Core \texttrademark~i7-8565U CPU @ 4.60 GHz x 4 and 16 GB of DDR4 RAM, Linux kernel 4.15.0-88-generic 64-bit version, Python version 3.8.10 and R version 4.0.3. The code for complexity and performance measurement as well as the data are available on GitHub\footnote{\url{https://github.com/MaxVidgof/process-complexity}}\footnote{\url{https://github.com/MaxVidgof/complexity-data}}.

\subsection{Forgetting} \label{sec:forgetting}
An important concern for prediction models is the potential evolution of the data-generation mechanism over time~\cite[p.525]{kuhn2013applied}. The available measures described in Section~\ref{sec:complexity}, such as sequence entropy and normalized sequence entropy, rely on counting the events distributed over different partitions of an automaton. However, they weight all events equally.
This can lead to undesirably high influence of older complex execution paths on current complexity measurements. 

Here, we consider the idea of \textit{forgetting}, which means the events that happen earlier should add less to the sequence entropy than more recent ones. To this end, 
we assign a weight to each event based on its timestamp, or, to be more precise, based on the time difference between each event and the most recent event in the log. Thus, the older the event, the more it will be discounted. 

There are two ways of doing so. The first, na\"ive way, is calculating the weight linearly as in Formula~\ref{eq:weight_linear}. Thus, this method is called \textit{linear forgetting}. \textit{Sequence entropy with linear forgetting} can be then computed similarly to the original \textit{sequence entropy} by summing up the weights of the events instead of counting them.

\begin{equation}
    \label{eq:weight_linear}
    w_l(e) = 1 - \frac{ts_{max} - ts(e)}{ts_{max}-ts_{min}}
\end{equation}

While linear forgetting provides a first glimpse of how forgetting can be incorporated into sequence entropy, it has a number of problems, all of which are connected to the weight assignment. First, the weight of the earliest observed event is 0, meaning the contribution of this event to process complexity is disregarded. This is an inadequate solution. Second, it implies a linear nature of forgetting itself, which does not reflect reality closely enough.

Thus, we introduce a more advanced method -- \textit{exponential forgetting}. It is similar to the first method, the only difference being a slightly more complex weighting Formula~\ref{eq:weight_exp}.

\begin{equation}
    \label{eq:weight_exp}
    w_e(e) = exp(-k\frac{ts_{max}-ts(e)}{ts_{max}-ts_{min}})
\end{equation}

With such weighting, the weight of the most recent event is 1 and earlier events have decreasing weights that never reach 0. In addition, the forgetting coefficient $k>0$ is introduced. It enables further control over the contribution of the older events. The larger the coefficient, the less the weight of the event. The coefficient is considered to be 1 by default and in this paper we proceed with this default value. If it is desired to decrease the weight of older events even more, a larger coefficient $k>1$ can be set. In the opposite case, one should use $0<k<1$.

\subsection{Data preparation}
Our dataset comprises 14 publicly available real-life event logs from Business Process Intelligence Challenge~(BPIC)~\cite{uuid:d9769f3d-0ab0-4fb8-803b-0d1120ffcf54,vanDongen_2015,uuid:5f3067df-f10b-45da-b98b-86ae4c7a310b,uuid:3301445f-95e8-4ff0-98a4-901f1f204972,uuid:d06aff4b-79f0-45e6-8ec8-e19730c248f1,uuid:52fb97d4-4588-43c9-9d04-3604d4613b51}. In order to use them for statistical analysis, we apply the following procedure. First, we split each event log into time periods. Then, we measure process performance and complexity for each period. Afterwards, we create a merged dataset with all event logs and add an \textit{industry} label to specify which industry the process belongs to. Finally, we remove the outliers. In this section, we describe these steps in more detail.

\textbf{Time periods}. We start by splitting the event logs into time periods. First, we extract the minimum and maximum timestamp of the log events. We then set the month of the earliest event to be the starting period and the month of the last event to be the end period. Afterwards, we split the event log into months using an \textit{intersecting} filter, as outlined in \cite{wurm2021measuring}. The filter assigns all traces to a period that started before or during the period and ended during or after the period. In other words, it selects all active traces in a given period. The choice of this filter further entails that a trace can be assigned to multiple periods.

Theoretically, it is possible to choose any granularity level at this point. I.e., we could choose shorter time periods like weeks or even days, but also longer ones like years. In any case, the choice of the the time interval mostly depends on case duration, e.g. setting time periods as granular as weeks if a process instance takes half a year on average will only increase the amount of data points without providing any additional value.

\textbf{Complexity measurement}. For each time period, we measure the complexity of the process, treating traces in every period as separate event logs. We use all metrics defined in Section~\ref{sec:complexity} and presented in Table~\ref{tab:measures}. Furthermore, we add forgetting to both simple and enriched sequence entropy, as outlined in Section~\ref{sec:forgetting}. Note that when calculating entropy metrics with forgetting, only a log partition (one month in our case) is considered, the minimal and maximal timestamp refer to the first and last event in this partition, not in the entire event log. It is also worth noting that from this point on we treat entropy metrics (both simple and enriched) as one group, which will be important in future steps.
In addition, we also measure a set of what we called \textit{generic} metrics. These are the metrics that can be measured out of the box by PM4Py\footnote{\url{https://pm4py.fit.fraunhofer.de/}} and include number of cases, number of activity repetitions, among others. We calculate a total of 38 complexity measurements for each time period.

\textbf{Performance measurement}. As already discussed, there are various ways to assess process performance, including time and cost dimensions. However, most publicly available event logs lack cost data, as is the case for the event logs we chose for analysis. We thus focus on measuring process time as an indicator for process performance. More specifically, we examine the throughput time of the respective processes. While using cycle time could potentially be more insightful, many event logs do not contain information about starting timestamps of activities, thus making it impossible to calculate cycle time. Throughput time, in contrast, can be easily calculated for any event log.
For each time period, we calculate the median throughput time of all traces in that period. While average throughput time might seem a more intuitive measure, median throughput time is more robust.

\textbf{Combined dataset}. After calculating the measurements for all logs, we combine them in a single dataset. In this dataset, each row consists of the originating event log, time period and corresponding measurements. 
We also want to control for industry-specific process characteristics that may influence performance but will not be captured by the complexity metrics.
Thus, we also introduce the variable \textit{industry} that specifies which industry the event log belongs to according to the SIC division~\footnote{\url{https://en.wikipedia.org/wiki/Standard\_Industrial\_Classification}}. 
Following the classification, we assigned BPIC~2011~\cite{uuid:d9769f3d-0ab0-4fb8-803b-0d1120ffcf54} to \textit{healthcare}, BPIC~2015~\cite{vanDongen_2015} and 2018~\cite{uuid:3301445f-95e8-4ff0-98a4-901f1f204972} to \textit{public administration}, BPIC~2017~\cite{uuid:5f3067df-f10b-45da-b98b-86ae4c7a310b} to \textit{finance}, BPIC~2019~\cite{uuid:d06aff4b-79f0-45e6-8ec8-e19730c248f1} to \textit{manufacturing}, and BPIC~2020~\cite{uuid:52fb97d4-4588-43c9-9d04-3604d4613b51} to \textit{education}.

\textbf{Outlier removal}. The last step of our data preparation procedure is the removal of outlier periods. At the beginning and at the end of each event log, there are periods that contain considerably less traces than the rest of the log. Our assumption is this is a by-product of data extraction. Consider the following case: if it is decided to extract all traces from January until December of year $Y$, then all the traces that were \textit{ongoing} in this time period will be extracted. However, some of them might have started earlier than January and some also ended later than December. It is indeed better to keep those traces in full rather than trim them (which might result in removing the start or end events and harm process discovery) or remove them entirely (in which case we would not have full information about resource usage). Still, in this case the extracted log will contain events occurring (at least) in years $Y-1$ and $Y+1$. 
For our approach, however, this is critical as this produces periods having not all traces, which, in turn, reduces the overall data quality. Thus, we filter out these periods either based on the event log description or based on the number of cases.
Note that we only remove outliers on the level of time periods, not individual traces.
The resulting dataset and some descriptive statistics are presented in Table~\ref{tab:dataset}.

\begin{table}[h]
\centering
\caption{Dataset description.}  
    \begin{tabular}{l|l|c|c|c|c|c}
    & & Event & & & & Median \\
    Source & Industry & logs & Periods & Traces & Events & throughput time\\
    \hline
    BPIC 2011~\cite{uuid:d9769f3d-0ab0-4fb8-803b-0d1120ffcf54} & Healthcare & 1 & 34 & 1143 & 150291 & 333 days \\
    BPIC 2015~\cite{vanDongen_2015} & Public adm. & 5 & 51 & 832-1409 & 44354-59681 & 38-108 days \\
    BPIC 2017~\cite{uuid:5f3067df-f10b-45da-b98b-86ae4c7a310b} & Finance & 1 & 13  & 31509 & 1202267 & 19 days\\
    BPIC 2018~\cite{uuid:3301445f-95e8-4ff0-98a4-901f1f204972} & Public adm. & 1 & 34 & 43809 & 2514266 & 267 days \\
    BPIC 2019~\cite{uuid:d06aff4b-79f0-45e6-8ec8-e19730c248f1} & Manufacturing & 1 & 13 & 251734 & 1595923 & 64 days \\
    BPIC 2020~\cite{uuid:52fb97d4-4588-43c9-9d04-3604d4613b51} & Education & 5 & 34 & 2099-10500 & 18246-86581 & 7-72 days  \\

    \end{tabular}
  \label{tab:dataset}%
\end{table}%

\subsection{Regression analysis}
After data preparation, we can now continue with building statistical models to explain throughput time based on process complexity. We start with two sets of independent variables. One with and one without industry as dummy variable.
For each set, we have the following procedure. First, we build the models automatically. Then, we reduce model size in terms of number of variables in two steps such that we are left with simple yet powerful models.
Note that we only consider linear combination of independent variables in this work.
The remainder of the section describes the procedure in more detail.

\textbf{Independent variables}. We use two sets of independent variables: complexity metrics with and without \textit{industry}. The reason for including \textit{industry} is to account for effects on throughput time that are in the nature of the specific process and do not depend on complexity of its execution sequences. On the other side, however, it is interesting whether process performance can be explained purely in terms of theory-backed complexity measures.

\textbf{Automated model selection}. We use automated model selection procedures to select the best regression models based on Akaike Information Criterion (AIC). We use three directions: \textit{forward}, \textit{backward} and \textit{both}. With \textit{forward} selection, we start with a small set of variables (only \textit{industry} if it is used or empty set of variables otherwise) and then add new variables one by one. At each step, the model with the lowest  AIC is selected for the further step. The procedure stops if adding more variables does not decrease AIC or if all variables are already included. In the \textit{backward} direction, we start from the model having all variables and remove them, also in a stepwise manner. Using \textit{both} directions, we start from a simple model and then at each step we can either add or remove a variable, depending on what yields the best AIC. 
As a result, we get 3 models for each of the 2 independent variable setups.

\textbf{Significant variables}. While the models produced in the previous step tend to have high explanatory power, they include a large number of independent variables, making them difficult to interpret and to use in practice. However, these models can often be further reduced in terms of the used variables. As a first step in this reduction, we remove all non-significant variables, i.e. variables with p-value larger than 0.001, from the models. We remove all variables that are not highly significant in one step. Interestingly, after this, some other variables in the model become less significant as well, thus we repeat the procedure until all the variables in the model are highly significant. In some cases, this procedure allows us to considerably reduce the size of the models, while keeping the explanatory power mostly unchanged. There are, however, cases, where the reduction leads to considerably lower explanatory power.

\textbf{Minimal models}. Finally, we create \textit{minimal} models. I.e., we further reduce the size of the models, such that at most one independent variable from each of the five categories (size, variation, distance, entropy, generic) is left. When selecting among variables in one category, the one with the lowest p-value is taken. In case two variables have the same p-value, the one that yields higher R-squared if left in the model is selected.

\section{Results}\label{sec:results}
In this section, we present our results. We start with automatically generated models that include \textit{industry} as dummy variable. We present summaries of full models, significant models as well as minimized models. Then, we show the best of the minimized models in terms of R-squared. Afterwards, we present models that use only \textit{theoretical} variables following the same structure.

\subsection{Full metrics}
We started with a model where throughput time depends on industry and automatically added complexity metrics. The best model was achieved with \textit{backward} selection. It included 23 variables and had R-squared of 0.9566887. Other models were very similar: \textit{forward} selection also produced a model with 23 variables  with very close R-squared of 0.9566887; selection in \textit{both} directions produced a slightly smaller model -- 18 variables -- with still similarly high R-squared of 0.9556369.

Restricting to only significant variables allowed to massively reduce the complexity of the models: to 11 for \textit{forward} and \textit{backward} selection and to 10 for selection in \textit{both} directions. The explanatory power of such models, however, only marginally decreased to roughly 0.94 for these models.

While being rather small, these models still contained some redundancy. They contained multiple variables belonging to the same categories of complexity defined in Section~\ref{sec:background}. In most cases, variables belonging to the same category were highly correlated, which is not surprising given they measure the same aspects of complexity. When removing such redundancy by leaving only one variable per category, we arrived at two models: the one resulting from minimizing the \textit{forward} selection model had 6 variables and R-squared of almost 0.92, and the one resulting from minimizing the \textit{backward} selection model had only 5 variables because the \textit{generic} variables were not among the significant variables, and had R-squared of 0.87. Minimizing the model resulting from selecting in \textit{both} directions resulted in the same model as from \textit{forward} selection and thus was dropped out.

The summary of the models can be viewed in Table~\ref{tab:industry}.
The best of the minimal models was the model resulting from \textit{forward} selection. It is presented in more detail in Table~\ref{tab:industry_best}. Note that while several possible values for the \textit{industry} variable are present (\textit{finance, healthcare, manufacturing} and \textit{public}), only one of them is present in the formula for each observation. The default value is the remaining industry, \textit{education}, thus in case of process in education no industry variable should be considered for the estimation.

\begin{table}[h]
\centering
\caption{Summary of regression models built from full set of metrics.}  
    \begin{tabular}{l|l|c|c}
    Size & Variable selection method & \ Number of variables   & R-squared \\
    \hline
    Full & forward        &   23      &   0.9554849 \\
    Full & backward        &   23      &   0.9566887 \\
    Full & both        &   18      &   0.9556369 \\
    \hline
    Significant & forward        &   11      & 0.9489275   \\
    Significant & backward & 11 & 0.9442622 \\
    Significant & both & 10 & 0.9391087 \\
    \hline
    Minimal & forward & 6 & 0.9195022 \\
    Minimal & backward & 5 & 0.8726173 \\
    
    \end{tabular}
  \label{tab:industry}%
\end{table}%

\begin{table}[h]
\centering
\caption{Best model with industry as dummy variable.}  
    \begin{tabular}{l|l}
    Variable & Estimate \\
    \hline
    Intercept &                               101.96221156   \\
Finance &                          -199.51131885   \\
Healthcare &                       280.14409159   \\
Manufacturing &                    -251.45111110   \\
Public &                         -124.96774620   \\
Magnitude &                                  0.00950598   \\
Level of detail &                             3.04980147   \\
Affinity &                                 -195.87822418   \\
Number of activity repetitions in period &   -0.00192884   \\
Enriched variant entropy &                                    -0.00079535   \\
    \hline
     R-Squared & 0.9195022 \\
    \end{tabular}
  \label{tab:industry_best}%
\end{table}%

\subsection{Theoretical metrics}
While the models presented above explain most of the variance in throughput time, they heavily rely on the \textit{industry} variable that accounts for industry-specific process characteristics. However, we see that relying only on \textit{theoretical} variables, i.e. only complexity variables, without any adjustments, still gives valuable results.

First, we can see that some of the automatically generated models are in fact even better than the ones shown in the previous section. Namely, the models achieved with \textit{forward} selection and selection in \textit{both} directions have slightly higher R-squared (0.969 vs 0.955) and at the same time have less variables (19 and 15 vs 23 and 18, respectively). The model generated with \textit{backward} selection is slightly worse, still on par with its counterpart.

Reducing the models to significant variables only gave differing results, some of them are very optimistic. Indeed, the model with \textit{forward} selection could be reduced to only 9 variables while still having R-squared of 0.94. For the other two models, the results are also very good, yet not that impressive. For instance, the model with \textit{backward} selection could be reduced from 26 to 18 variables while maintaining its R-squared of over 0.94, it is still very large. The model with selection in \textit{both} directions could be reduced to 7 variables only, however, at a price of considerably lower R-squared of 0.82.

The models could be reduced even further, with the smallest minimal model containing only 2 variables. However, the explanatory power of such models barely reaches 0.8 in the best case. The models are summarized in Table~\ref{tab:theoretical}. The best minimal model containing only theoretical variables is presented in more detail in Table~\ref{tab:theoretical_best}. Interestingly, the significant model for \textit{forward} selection only contained complexity metrics from 2 categories, thus the minimal model has only 2 variables.

\begin{table}[h]
\centering
\caption{Summary of regression models built from theoretical metrics.}  
    \begin{tabular}{l|l|c|c}
    Size & Variable selection method & \ Number of variables   & R-squared \\
    \hline
    Full & forward        &   19 & 0.9696743 \\
    Full & backward        &   26 & 0.9513178 \\
    Full & both        &   15 & 0.9698724 \\
    \hline
    Significant & forward        &   9 & 0.93589 \\
    Significant & backward & 18 & 0.9479194 \\
    Significant & both & 7 & 0.8202257 \\
    \hline
    Minimal & forward & 2 & 0.7965621 \\
    Minimal & backward & 4 & 0.6385549 \\
    Minimal & both & 4 & 0.7294595 \\
    
    \end{tabular}
  \label{tab:theoretical}%
\end{table}%

\begin{table}[h]
\centering
\caption{Best model with theoretical variables only.}  
    \begin{tabular}{l|l}
    Variable & Estimate \\
    \hline
    (Intercept) & -142.1478 \\
    Average trace length & 4.8333 \\
    Affinity & 274.9644 \\
    \hline
    R-Squared & 0.7965621 \\
    
    \end{tabular}
  \label{tab:theoretical_best}%
\end{table}%

\section{Discussion}\label{sec:discussion}

\subsection{Implications}
During this work, we have made some interesting observations. First of all, we see that
industry alone explains 80\% of variance in the dependent variable throughput time. 
This means that processes in different industries and in different companies are so different in their nature that knowing where the process is executed allows us to infer a lot about its throughput time without considering its complexity at all.

However, adding the complexity dimension on top of the industry allows us to gain even more insightful information, explaining up to 95.6\% of variation in throughput time. One can look at it from different perspectives. On the one hand, when having 80\% of the output explained by one categorical variable that does not even need any further computation, one can say that all possible additions to it are only marginal and are not worth considering. On the other hand, being able to explain more than 95\% of variance in throughput time is a valuable capability that is worth the effort. In addition, there is a compromise solution with the minimal models. One can still achieve remarkable explanatory power using only a handful of metrics: 5 complexity metrics on top of industry can explain roughly 92\% of variance.

Despite having such high explanatory power, models containing \textit{industry} as an independent variable have received criticism as the observations used to build them only consider one or two processes per industry and thus are not necessarily representative. In the light of this criticism, we also developed models explaining throughput time solely by the complexity of the corresponding processes. The good news is that these \textit{theoretical} variables successfully managed to compensate the information gained by industry. After all, the industry a process belongs to is not something that directly influences process performance by itself but rather a factor that contributes to how the process is set up, thus it is not surprising that these differences are (at least to some extent) visible in the complexity of the process.

These models with theoretical variables achieve similar results in terms of R-squared while having slightly smaller variable counts. Interestingly, the full models that were generated in the first step have even slightly higher R-squared. This resulted from different starting points: for the models with industry, the lower boundary for model selection was a model already containing the industry because it was considered a baseline. The theoretical models, instead, had a constant as their lower boundary. The interpretation of this is that if we set no boundaries and allow (but do not force) selecting the \textit{industry} variable, it might be the case that the best models will still not contain it, which even better supports our idea of being able to explain process performance using its complexity only.

It is also interesting to look at these models in more details because they are structurally different from the ones including \textit{industry}. In the first step, these two kinds of models are similar in terms of both the number of variables and R-squared.
In the second step, where we restrict the models to only containing significant variables, some differences become visible. Theoretical models achieved with \textit{forward} selection and selection in \textit{both} directions have slightly smaller R-squared than their counterparts including \textit{industry}. However, this can be attributed to just having also slightly smaller number of variables. The model achieved with \textit{backward} selection, however, does not fit the pattern. Its full version had a lot of significant variables, thus it could not be reduced much, which also allowed it to keep most of its R-squared. 
The most interesting differences become visible in the last reduction step, where we only choose one variable per category. Theoretical models contain a more homogeneous sets of variables, i.e. more variables from one category, while some categories can be missing entirely. Thus, reducing them to minimal models yields much smaller (2-4 variables) but also much less powerful (R-squared 0.63-0.79) models. Note that models containing industry cannot have such low values at all as \textit{industry} alone would have R-squared of 0.8.

Up to this point, our goal was to develop the most simple yet powerful explanatory models. However, as we achieved this, we asked ourselves whether we can tweak the models a bit further to gain more explanatory power while not increasing the complexity of the models by much. The first approach we tried was to add interplay between the variables in our models. This, however, was not very fruitful. Adding pairwise interplay terms between all variables in models did not improve R-squared. Adding all possible interplay terms between two but also more variables in the model did increase the R-squared (for instance, we achieved R-squared of 0.96 with only 6 variables for the model including industry), however, such terms are very difficult to explain.

Another approach that we took was clustering the event logs based on their median throughput times and developing separate models for each cluster. With this approach, we could achieve slightly better (or smaller) models in the first step. However, R-squared falls drastically when we try to minimize models. 

The last observation that we did, also going in the direction of clustering, is that in the end processes are different and while we can explain a lot of their variance in terms of complexity, there is no one-fits-all solution. 
Our results and models should be thus considered as toolbox, and the practitioners should analyze which exact variables makes sense in case of their processes.

\subsection{Future Work}
We see several promising avenues for future research. First, our quantitative analysis of throughput time can be extended in several directions. On the one hand side, future work can use principal component analysis to cluster the independent variables. This would not only reduce model size, but might lead to interesting insights how different theoretical variables can be empirically grouped together. On the other hand side, our analysis presented in this paper can be complemented by the prediction of throughput time. We deem our models a suitable starting point for such an endeavour. 

Second, behavioral studies can investigate how process complexity develops over time. For example, how process complexity is reduced and increases in course of business process standardization initiatives. Such a study could focus on the specific actions taken by management and unpack how they influence process performance and overall complexity. We deem such studies particularly fruitful, if they can complement insights from event logs with detailed interviews with key stakeholders, such as managers and process experts.  

Third, more generally, there are plenty of opportunities for behavioral business process research due to the increasing availability of digital trace data \cite{DBLP:conf/hicss/GrisoldWMB20}. Future research can make use of digital trace data from event logs to investigate how business process change over time, contributing to theory on business process change and routine dynamics \cite{mendling2021philosopher,wurm2021business}.

\section{Conclusion}\label{sec:conclusion}
In this paper, we reported on a study in which we empirically examined the link between process complexity and throughput time. Based on 14 event logs and 38 different process complexity metrics, we created various statistical models that explain the throughput time of business processes. Our models are able to explain a large share of the variance in the throughput time, reaching R-squared values of up to 0.96. Our results provide important implications for research on process complexity and process standardization. Practitioners can use our implementation of the different complexity measures to monitor their processes. 

%
%

\bibliographystyle{splncs04}
\bibliography{bib}

\end{document}